\documentclass[twocolumn,showpacs,amsmath,amssymb]{revtex4}
\usepackage{epsfig}
\usepackage{amsmath}
\usepackage{amssymb}
\usepackage{graphics}
\usepackage{color}
\usepackage[colorlinks=true,citecolor=blue,linkcolor=blue]{hyperref}

\DeclareMathAlphabet{\bi}{OML}{cmm}{b}{it}

\begin{document}
\title{Thermoelectric power factor in nanostructured materials with randomized nanoinclusions}
\author{Vassilios Vargiamidis, Samuel Foster, and Neophytos Neophytou}
\address{School of Engineering, University of Warwick, Coventry, CV4 7AL, UK}%
\begin{abstract}
We investigate the electric and thermoelectric transport coefficients of nanocomposites using the Non-Equilibrium Green's Function (NEGF) method, which can accurately capture the details of geometry and disorder in these structures. We consider here two-dimensional (2D) channels with embedded nanoinclusions (NIs) modelled as potential barriers of cylindrical shape and height $V_B$. We investigate the effect of randomness of the NIs on the thermoelectric power factor by varying the positions, diameter, and heights of the barriers according to a Gaussian probability distribution. We find that the power factor shows indications of tolerance to variations in the parameters of the NIs when the Fermi level is placed into the bands and $V_B \sim k_B T$. These results could  be experimentally relevant in the design of nanocomposites for thermoelectric applications.
\end{abstract}
 \maketitle

\section{Introduction}

Thermoelectric energy conversion is the ability of a device to convert a temperature gradient into an electrical current. The efficiency of the thermoelectric conversion is quantified by the dimensionless parameter, called figure of merit
\begin{equation}
ZT = \frac{\sigma S^2 T}{\kappa} ,
\label{eq1}%
\end{equation}
where $\sigma$ is the electrical conductivity, $S$ is the Seebeck coefficient, $T$ is the absolute temperature, and $\kappa = \kappa_e + \kappa_p$ is the total thermal conductivity that is usually split into electron $(\kappa_e)$ and phonon $(\kappa_p)$ contributions. The quantity $\sigma S^2$ is known as the power factor (PF).

During the last few years the quest for a highly efficient thermoelectric device has attracted considerable attention due to potential technological and industrial applications \cite{vin09,neo15,zia16}. Traditionally the figure of merit has been constrained to values $ZT \approx 1$. The optimization of the figure of merit has proved to be a quite challenging problem. This is due to the fact that the quantities that control it are inversely related, and changing one of them leads to the adverse change of the other. However, significant progress has been made recently towards improving the value of $ZT$; namely, it has been demonstrated that $ZT \approx 2.2$ in some nanostructures \cite{wu1,zhao1}, which is mainly due to the large reduction in their thermal conductivities compared to bulk material values.

Nanostructuring has been a technique largely employed in achieving performance improvements in various settings. For example, it has been shown that dense dislocation arrays formed at low-energy grain boundaries can lead to a reduction of thermal conductivity \cite{wang15} giving rise to improved $ZT$. Suppression of thermal conductivity has also been observed when small amounts of $CuBr_2$ are incorporated in $Bi_2S_3$ thermoelectric materials \cite{meng17}, in the case of nanometer sized inclusions of $Sb$, $Bi$, and $InSb$ in bulk $PbTe$ \cite{soot06}, and in $\alpha-MgAgSb$-based materials \cite{pei15} due to its beneficial lattice dynamics properties and the multiscale microstructure. In addition, reduction of the thermal conductivity via dense dislocation scattering along the grain boundaries formed for $Yb$-filled $CoSb_3$ skutterudites has been observed \cite{mao18}, and also in nano-micro-porous skutterudites \cite{koba17}.

There are several other methods for reducing $\kappa$; namely, by using superlattices \cite{mizuno1}, heavy doping \cite{ikeda1}, nanoporous materials \cite{verdier1}, and boundary scattering in low-dimensional materials \cite{hoch08}, to name a few. However, one of the most successful methods is the use of nanoinclusions (NIs) \cite{biswas1,gayner1,zou1}, which cause scattering of high-energy phonons resulting in a significant reduction in $\kappa$. In fact, by embedding NIs within PbTe in a hierarchical manner \cite{biswas1,herem1}, record high $ZT = 2.2$ was achieved due to drastic reduction in $\kappa$.
\begin{figure}[t]
\vspace{0.25in}
\includegraphics[width=7.4cm,height=5.0cm]{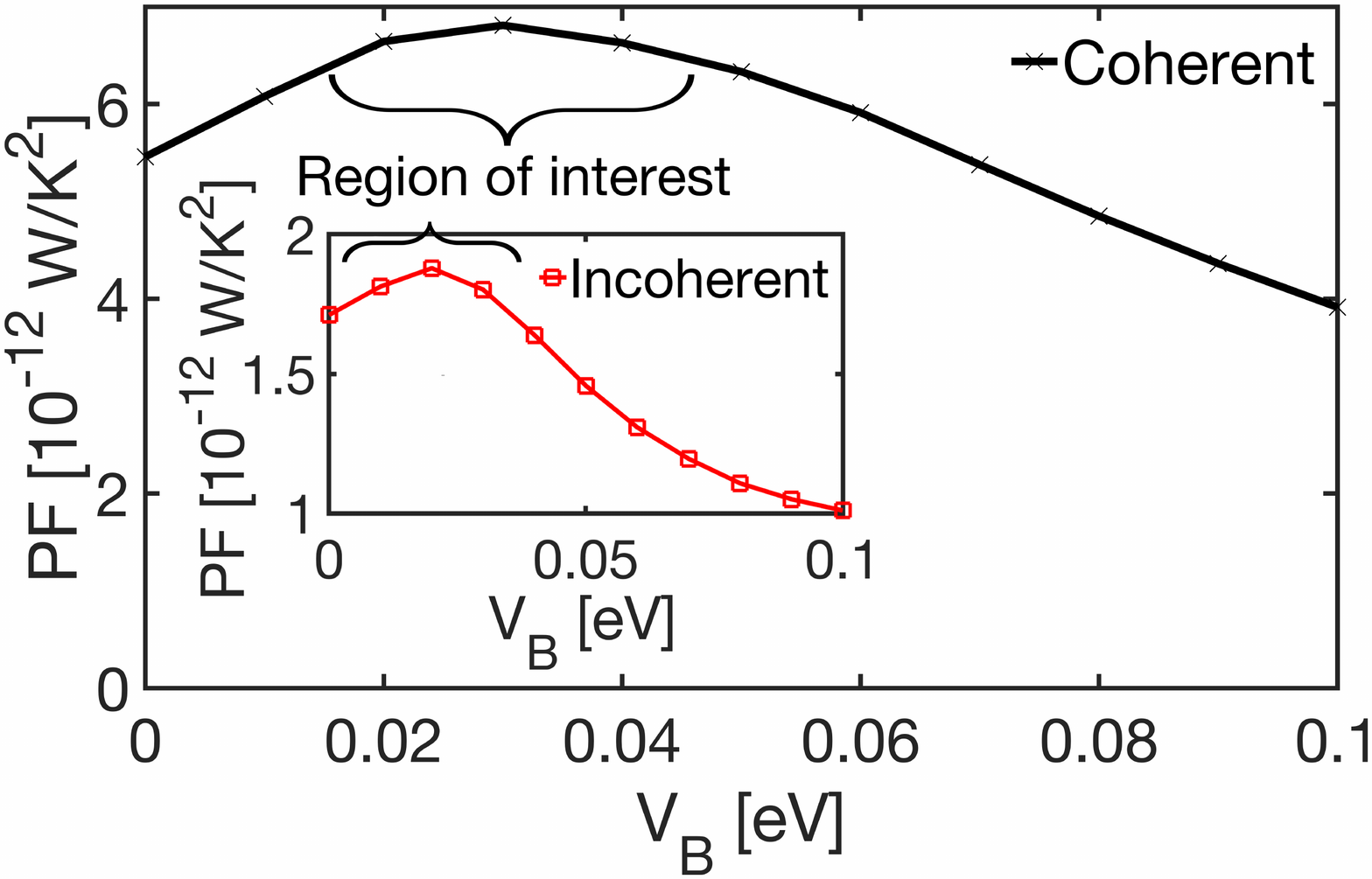}
\vspace*{-0.15cm} \caption{\label{fig1} (Colour online)
The PF of a matrix material containing an $8\times4$ hexagonal array of nanoinclusions vs nanoinclusion barrier height for two scattering regimes: Ballistic (coherent) transport (black-cross line), and acoustic phonon scattering transport (red-square line). With the brackets we indicate the barrier height range of interest. Results are taken from Ref.~\cite{fos1}.
}
\end{figure}

Even though the effects of NIs on the thermal conductivity have been investigated to some extent \cite{verdier1} their influence on the PF has not been as clearly determined \cite{zou1,liu1,zou2} and the effects of random variations of the NIs on the PF have not been taken fully into account. It becomes desirable therefore to give a detailed account of the effects of random variations of the NIs on the PF.

We have previously performed studies on the PF of such types of nanostructures \cite{fos1}, but we only considered NIs placed in idealized ordered positions. In that work we speculated that such structures exhibit robustness to structural and geometrical variations. In particular, under highly degenerate conditions (i.e., the Fermi level $E_F$ placed well into the bands), we showed that when placing the barrier heights of the NIs $V_B \sim k_B T$ above the band edge, i) the PF peaks, and ii) the effects of variations in the barrier height and NI density on the PF are significantly suppressed (see Figs.~3 and 5 in Ref. \cite{fos1}). In Fig.~1 we show the PF versus NI barrier height of an $8\times4$ hexagonal array of NIs in a 2D channel matrix (as shown in Fig.~2(a)) for two different scattering regimes; namely, ballistic (black-cross line) and acoustic phonon scattering (red-square line) regime. [Data in this figure is taken from Ref.~\cite{fos1}]. With the brackets we indicate the regions of interest, i.e. when $V_B \sim k_B T$ above the band edge. In these regions the PF shows only a minor change as the barrier height is varied.
\begin{figure}[t]
\vspace{0.0in}
\includegraphics[width=6.7cm,height=9.0cm]{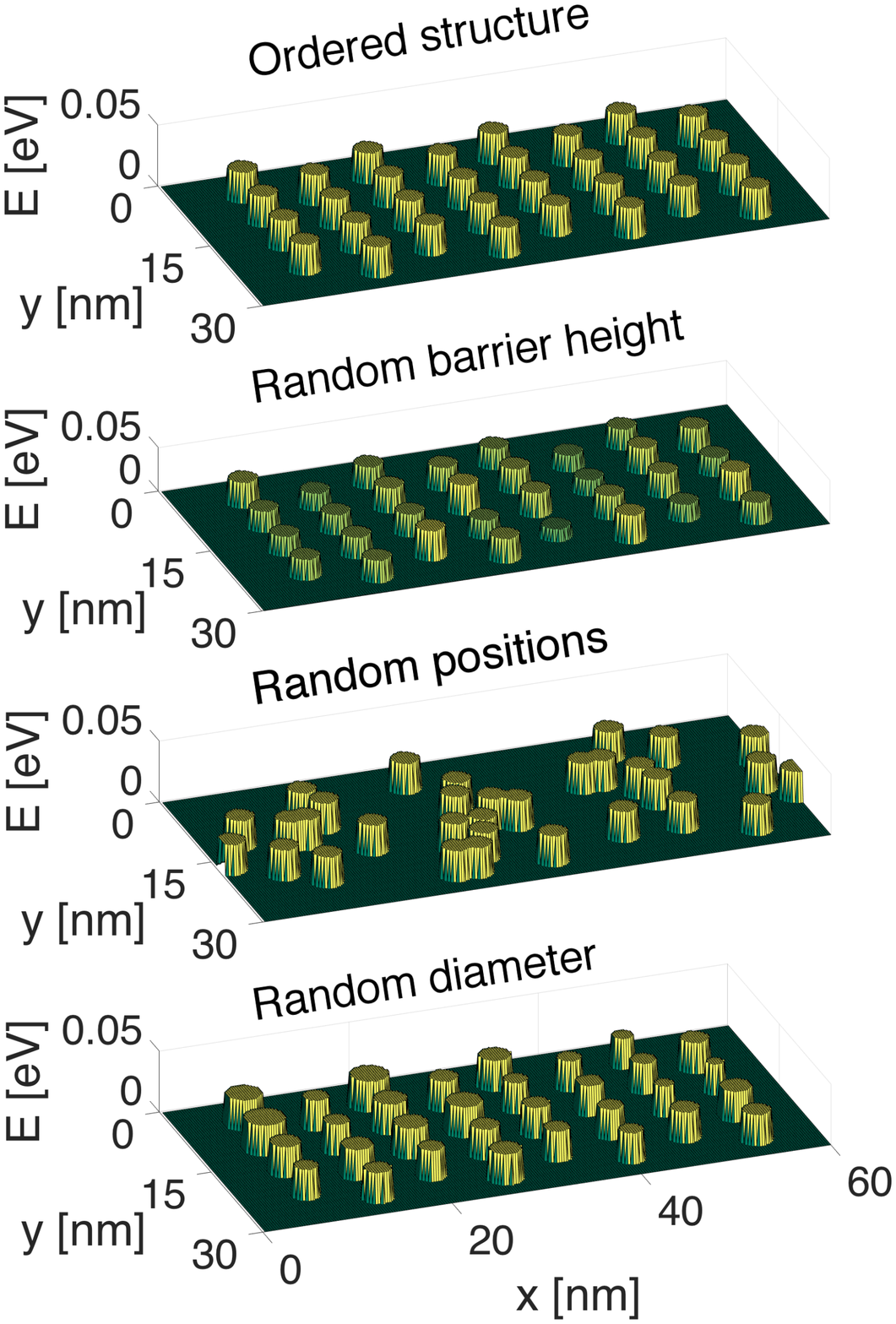}
\vspace*{0.05cm} \caption{\label{fig2} (Colour online)
The materials with embedded NIs geometries we consider: (a) The ordered structure consisting of $8\times4$ hexagonal array of NIs of barrier height $V_B = 0.05$ eV and diameter $d = 3$ nm. (b) An $8\times4$ hexagonal array of NIs with random barrier height and diameter $d = 3$ nm. (c) A random distribution of $32$ NIs of barrier height $V_B = 0.05$ eV and diameter $d = 3$ nm. (d) An $8\times4$ hexagonal array of NIs of barrier height $V_B = 0.05$ eV and random diameter. 
}
\end{figure}

The purpose of this paper is to extend our previous work \cite{fos1} aiming to not only verify the tolerance of the PF to such types of variations and randomness in nanocomposite materials, but also demonstrate the suitability of the Non-Equilibrium Green's Function (NEGF) method to perform such simulations, where the geometry can take random forms, and transport can vary from the ballistic to the diffusive regime. The NIs are modelled as potential barriers of cylindrical shape (see Fig.~2). In particular, we investigate the effect of randomness of the NIs on the PF; namely, the positions, diameter, and heights of the barriers are varied according to a Gaussian probability distribution. Such random variations, which reflect the imperfections in nanocomposite materials, are inherently present in any real system and their effects have to be taken into account. We find that materials with NIs, once doped in highly degenerate levels and the barriers are one or two $k_B T$ above the band edge, are robust to variations in the NI barrier height, the NI diameter, and their geometry. Our results could prove useful in the design of nanocomposite materials that provide PF robustness.

The rest of the paper is organized as follows. In Sec.~II, we briefly describe the NEGF method and the computational scheme. In Sec.~III we present the results of our analysis and we conclude with a summary in Sec.~IV.

\section{NEGF Method and computational scheme}

In order to calculate electronic transport properties we employ the NEGF method \cite{datta05}. The formalism of NEGF captures all quantum mechanical effects and is particularly suitable for the treatment of electron-phonon (e-ph) interactions in nanoscale materials. We briefly outline the formalism of the NEGF method and describe the computational scheme that we used to simulate electronic transport in the 2D channel.

In the NEGF method a system/device, described by a Hamiltonian $H$, is connected to two contacts (left and right) which are represented by self-energy functions $\Sigma_L$ and $\Sigma_R$. These self-energies represent the influence of the semi-infinite Left and Right leads on the device, respectively. Note that $\Sigma_L$ and $\Sigma_R$ are energy dependent, and non-Hermitian. The e-ph scattering process in the device enters the NEGF formalism through the self-energy function $\Sigma_S$. One can view the scattering process as just another contact described by $\Sigma_S$, similar to the actual contacts described by $\Sigma_L$ and $\Sigma_R$.

The retarded Green's function for the device is given by \cite{datta05}
\begin{equation}
G(E) = \left[ \left( E + i \eta^+ \right) I - H - \Sigma(E) \right]^{-1} ,
\label{eq2}%
\end{equation}
where $\eta^+$ is an infinitesimally small positive number which pushes the poles of $G$ to the lower half plane in complex energy, $I$ is the identity matrix, and $\Sigma(E)$ is the sum of the self-energies
\begin{equation}
\Sigma(E) = \Sigma_L(E) + \Sigma_R(E) + \Sigma_S(E) .
\label{eq3}%
\end{equation}
For ballistic transport $\Sigma_S(E) = 0$. The self-energy terms have two effects. One is to change the eigenstates and shift the eigenenergies of the Hamiltonian $H$. The second effect is to introduce an imaginary part to the energy which is determined by the broadening functions $\Gamma_1$, $\Gamma_2$, and $\Gamma_S$ defined by
\begin{equation}
\Gamma_{L, R, S} (E) = i \left[ \Sigma_{L, R, S}(E) - \Sigma_{L, R, S}(E)^\dagger \right] ,
\label{eq4}%
\end{equation}
where $\Sigma_{L, R, S}(E)^\dagger$ represents the Hermitian conjugate of $\Sigma_{L, R, S}(E)$. These broadening functions determine the escape rate of an electron initially placed in an energy level of the device into the left and right leads, i.e., the imaginary part of $\Sigma_{L, R}(E)$ multiplied by $-2$ is the scattering rate of electrons to the Left or Right lead $\left( \Gamma_{L, R}(E) = - 2 \text{Im} \left[\Sigma_{L, R}(E) \right] \right)$. It proves useful and convenient to define the in-scattering self-energies due to contacts as
\begin{equation}
\Sigma_{L, R}^{in} (E) = - 2 \text{Im}  \left[ \Sigma_{L, R} (E) \right] f_{L, R} (E) ,
\label{eq5}%
\end{equation}
where $\text{Im}[...]$ is the imaginary part and $f_{L, R}$ is the Fermi distribution for the left and right leads. These self-energies physically represent in-scattering of electrons from the semi-infinite leads to the device. They depend on the Fermi factor in the contacts, $f_L$ and $f_R$, and the strength of coupling between contacts and device, $\text{Im} \left[ \Sigma_L(E) \right]$ and $\text{Im} \left[ \Sigma_R(E) \right]$. Similarly, the out-scattering self-energies are defined as
\begin{equation}
\Sigma_{L, R}^{out} (E) = - 2 \text{Im}  \left[ \Sigma_{L, R} (E) \right]  \left[ 1 - f_{L, R} (E) \right] .
\label{eq6}%
\end{equation}
Note that $\Sigma_{L, R}^{out} (E)$ is similar to $\Sigma_{L, R}^{in} (E)$ apart from the fact that the probability of finding an occupied state in the contact, $f_{L, R}$, is replaced by the probability of finding an unoccupied state in the contact, $1 - f_{L, R}$. With $\Sigma_{L, R}^{in/out} (E)$ one can express the electron and hole correlation functions as
\begin{equation}
G^n (E) = G(E) \Sigma_{L, R}^{in} (E) G^\dagger (E) ,
\label{eq7}%
\end{equation}
\begin{equation}
G^p (E) = G(E) \Sigma_{L, R}^{out} (E) G^\dagger (E) .
\label{eq8}%
\end{equation}

Assuming that the system is described by a set of one-dimensional grid/lattice points with uniform spacing $a$, and making the nearest neighbour tight binding approximation, the current density between grid points $q$ and $q+1$ is given by
\begin{eqnarray}
\nonumber J_{q, q+1} = \frac{i e}{\hbar}
\\* &&\hspace*{-1in} \times (2)  \int_{-\infty}^{\infty} \frac{d E}{2 \pi} \left[ H_{q+1,q} G_{q, q+1}^n (E) - H_{q,q+1} G_{q+1, q}^n (E) \right]   ,
\label{eq9}%
\end{eqnarray}
where $H_{q+1, q} = H_{q, q+1}^\dagger$ are the hopping matrix elements of the Hamiltonian, and $(2)$ is for the two spin directions. The effective mass is taken to be $m=1$ throughout the whole material and in the NIs.
\begin{figure}[t]
\vspace{0.0in}
\includegraphics[width=5.4cm,height=7.7cm]{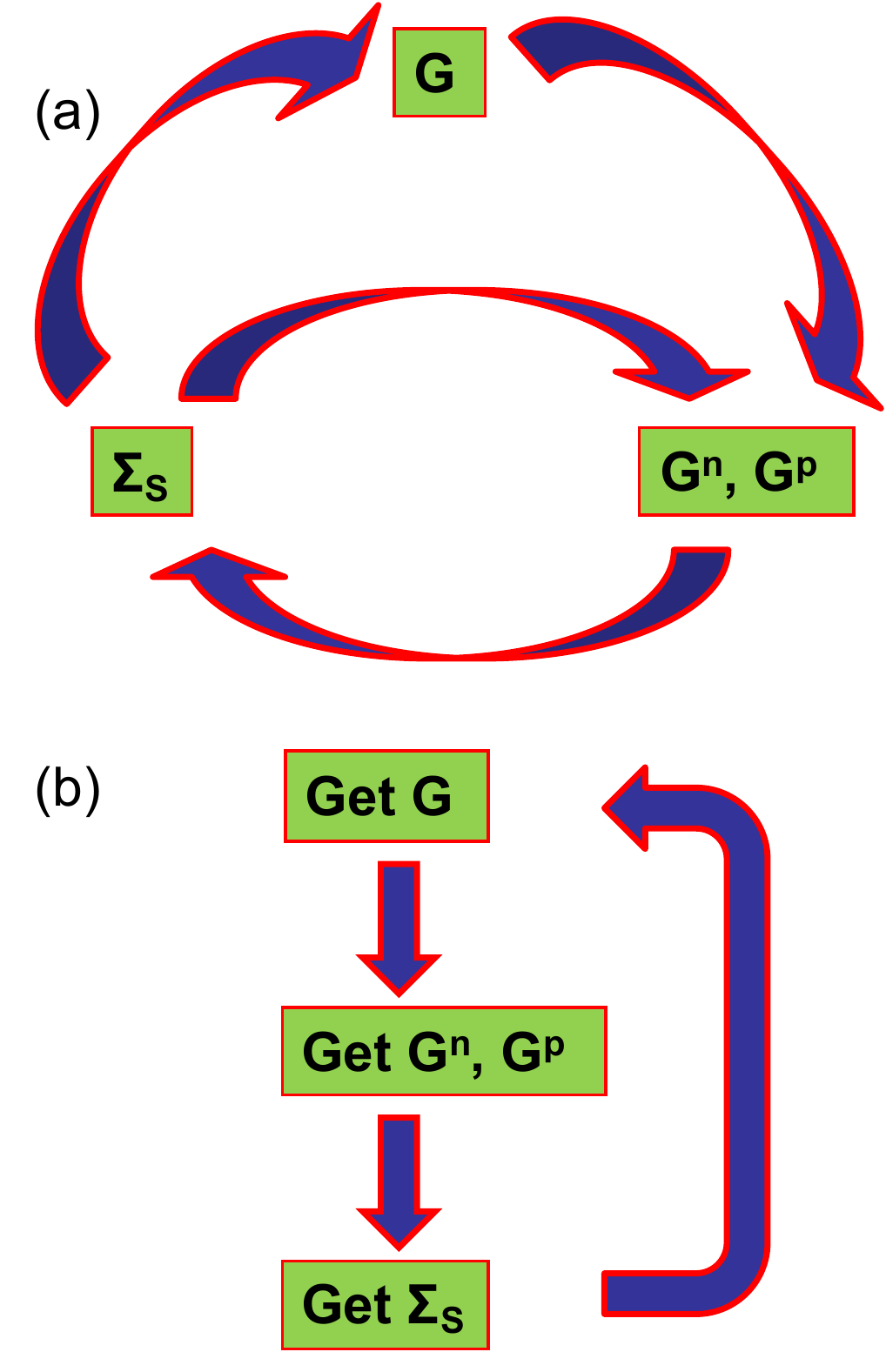}
\vspace*{0.05cm} \caption{\label{fig2} (Colour online)
Computational scheme for the self-consistent calculation
}
\end{figure} 

A second source for in-scattering and out-scattering of electrons from an occupied state is the e-ph interaction. The self-energy at point $q$ and energy $E$ has two terms corresponding to scattering from $(q, E+\hbar \omega_{ph})$ and $(q, E-\hbar \omega_{ph})$. Within the Born approximation the in-scattering self-energies into a fully empty state is \cite{mahan87}
\begin{eqnarray}
\nonumber \Sigma_{S_{q}}^{in} (E) = \sum_{\eta} D_{q}^\eta [ n_B(\hbar \omega_{ph}) G_q^{n} (E-\hbar \omega_{ph}) 
\\* &&\hspace*{-2.25in} + (n_B ( \hbar \omega_{ph}) + 1) G_{q}^{n} ( E + \hbar \omega_{ph}) ]    ,
\label{eq10}%
\end{eqnarray}
where $D_{q}^\eta$ represents the e-ph scattering strength at grid point $q$, $\eta$ is the phonon subband, $n_B$ is the Bose-Einstein distribution function for phonons of energy $\hbar \omega_{ph}$, and $G_q^{n} (E-\hbar \omega_{ph})$ is the electron density at $E-\hbar \omega_{ph}$.  The first and second terms in Eq.~(\ref{eq10}) represent in-scattering of electrons from $E - \hbar \omega_{ph}$ (phonon absorption) and $E + \hbar \omega_{ph}$ (phonon emission) to $E$, respectively. The out-scattering self energy, $\Sigma_{S_{q}}^{out} (E)$, from a fully filled state is given as \cite{mahan87} 
\begin{eqnarray}
\nonumber \Sigma_{S_{q}}^{out} (E) = \sum_{\eta} D_{q}^\eta [ (n_B(\hbar \omega_{ph}) + 1) G_q^{p} (E-\hbar \omega_{ph}) 
\\* &&\hspace*{-2.25in} + n_B ( \hbar \omega_{ph})  G_{q}^{p} ( E + \hbar \omega_{ph}) ]    ,
\label{eq11}%
\end{eqnarray}
where $G_q^{p} (E-\hbar \omega_{ph}) $ and $G_q^{p} (E+\hbar \omega_{ph})$ are the densities of unoccupied states at $E-\hbar \omega_{ph}$ and $E+\hbar \omega_{ph}$. The first and second terms in Eq.~(\ref{eq11}) represent out-scattering of electrons from $E$ to $E+\hbar \omega_{ph}$ (phonon emission) and $E-\hbar \omega_{ph}$ (phonon absorption), respectively. However, in this work we consider only acoustic phonon scattering where there is no change in energy.

The parameter $D_{q}^\eta$ is related to the phonon deformation potential $D_{op}$ by
$D_{q}^\eta = \hbar D_{op}^2 F / 2 \rho \omega_o a^3$, where $F$ is the wave function overlap, $\rho$ is the mass density, and $\omega_0$ is the optical phonon frequency \cite{KimJAP12,kos07}. In this paper we consider a constant e-ph scattering strength, i.e., $D_{q}^\eta = D_0$, which is the same throughout the channel. The value is chosen such that the mean-free path of the system is 15 nm (comparable to common semiconductors such as silicon \cite{neoprb11}) using the method described in Ref. \cite{thesberg2}. That is, the phonon strength is chosen such that the ballistic conductance drops to half in channels of length $15$ nm. The appropriate value of $D_0$ was found to be $D_0 = 0.0026 eV^2$.

In Fig.~3 we show the computational scheme that we used to perform the numerical simulations. One can see that in order to compute the the current we need $G^n(E)$. However, we note from Eq.~(\ref{eq7}) that $G^n(E)$ depends on the Green's function $G(E)$ and the scattering self-energies. The scattering self-energies on the other hand depend on $G^n(E)$, and $G(E)$ depends on the scattering self-energies. These interelations are shown in Fig.~3(a). Thus, in order to calculate the current through the material we perform self-consistent calculations as shown in Fig.~3(b), where we provide an initial value for $G(E)$, $G^n(E)$, $G^p(E)$, and $\Sigma(E)$ (in this case the ballistic quantities that can be computed without self-consistency) and then loop over them by turning on phonon scattering until convergence is accomplished. The convergence criteria for the ensuing self-consistent calculation is chosen to be current conservation; namely, we consider that convergence is achieved when the current is conserved along the length of the channel to within $1\%$.

We note here that other methods can be employed to describe transport in complex, large-scale disordered materials, each with its own approximations and computational difficulties. The NEGF method is a fully quantum mechanical method that captures all basic physics at the microscopic level, and gives energy resolved information, however it can become computationally expensive. On the other hand, another method, the Wigner function, captures the potential variation and its effect on distribution of carriers across the material, and can also model time dependent transport. It gives an alternative perspective on quantum transport and is less computationally intensive than NEGF. However, the dissipator term can be difficult to treat sometimes, and is often necessary modelled using a Boltzmann collision operator \cite{jung09}. A coupling scheme of NEGF and Wigner function approach has also been explored \cite{baum08}. Monte Carlo methods solve the Boltzmann transport Equation statistically, and their path tracing nature allows for much larger simulation domains, however, quantum effects are absent.

\section{Results}

The 2D channel that we consider is of length $L = 60$ nm and width $W = 30$ nm (see Fig.~2).The e-ph scattering strength is set at $D_0 = 0.0026$ eV$^2$. The conduction band is set at $E_c = 0$ eV and the Fermi level is placed at $E_F = 0.05$ eV. 
\begin{figure}[t]
\vspace*{-0in}
\hspace*{-0.24in}
\includegraphics[width=9.6cm,height=13cm]{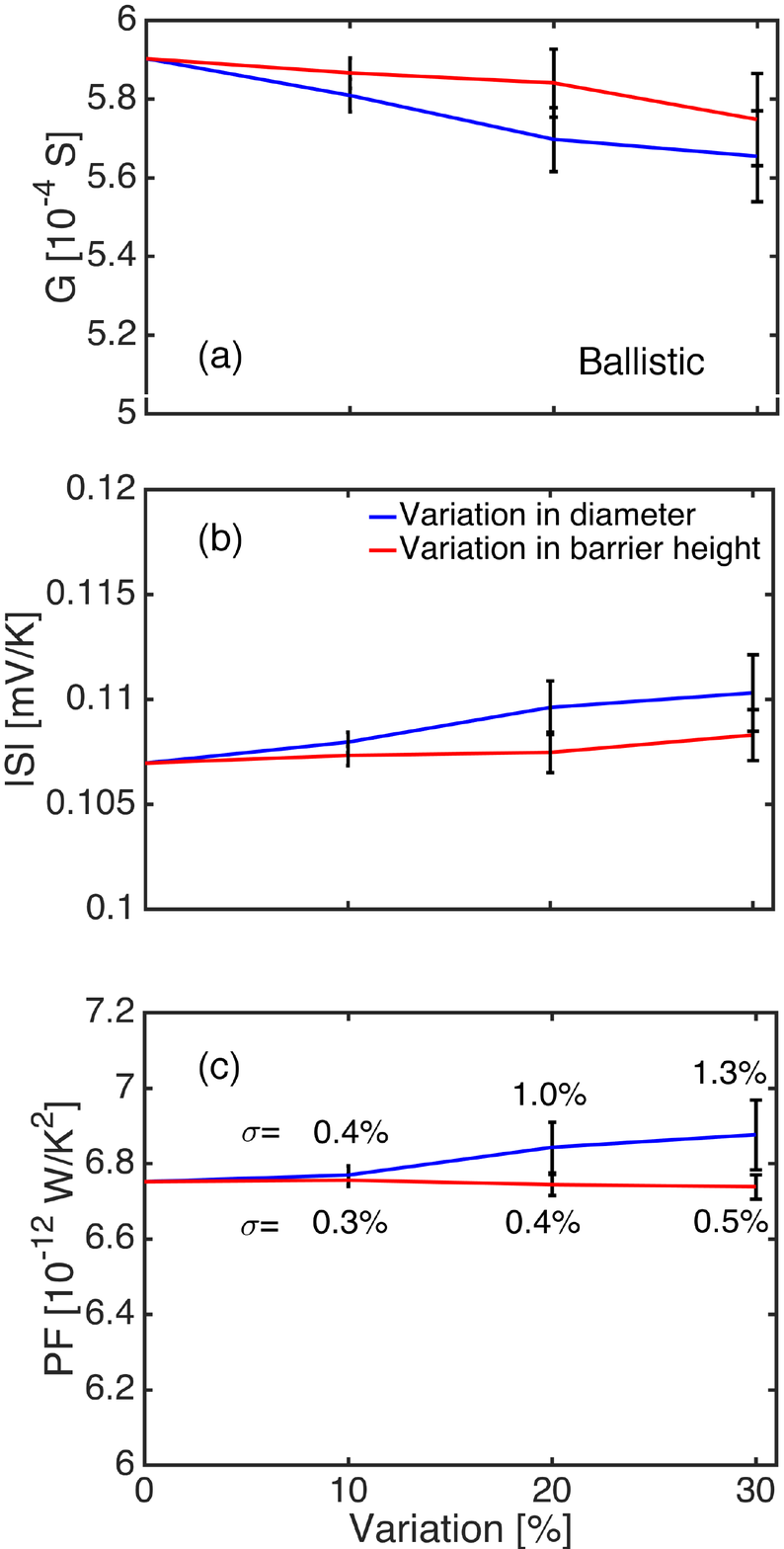}
\vspace*{0.05cm} \caption{\label{fig3} (Colour online)
The effect of variations of the NI diameter and barrier height on the thermoelectric coefficients. (a) The electrical conductance, (b) the Seebeck coefficient, (c) the power factor, versus the percentage variation from the nominal values. Variations in the radius (blue lines) and barrier height (red lines) are shown. Each data point is the average of at least 10 simulations and the error bars indicate the standard deviation of the results (shown by the labels). 
}
\end{figure}
\begin{figure}[t]
\vspace{-0.07in}
\hspace*{-0.6in}
\includegraphics[width=11.5cm,height=14.5cm]{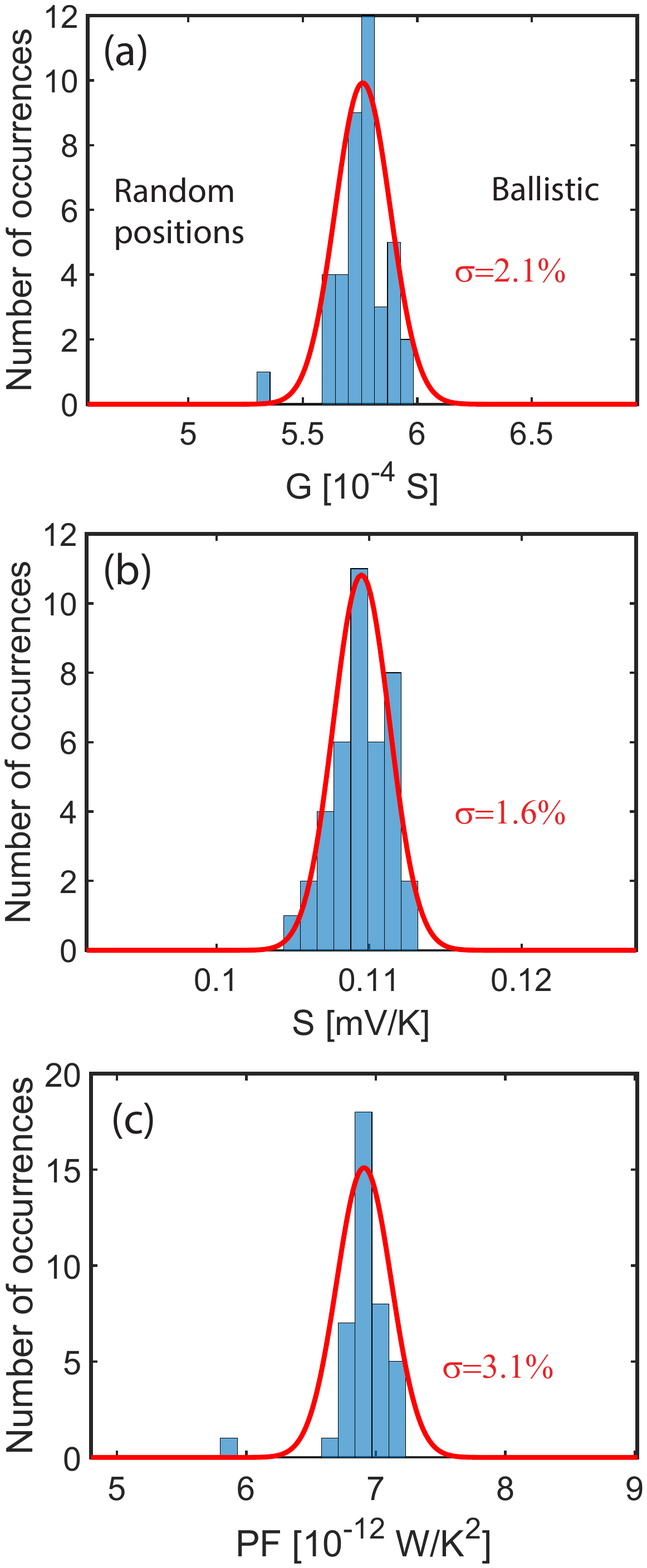}
\vspace*{0.05cm} \caption{\label{fig4} (Colour online)
Histograms of the values of (a) the conductance $G$, (b) the Seebeck coefficient $S$, (c) the PF $GS^2$ for randomised geometries with randomized pore positions under ballistic transport conditions.
}
\end{figure}

We begin with an investigation into the effect of variations in a ballistic channel. In Fig.~4 we show the thermoelectric coefficients, conductance $G$ (Fig.~4(a)), Seebeck coefficient $S$ (Fig.~4(b)), and PF $GS^2$ (Fig.~4(c)) of a ballistic channel with an $8 \times 4$ hexagonal array of NIs as two different parameters are varied: the barrier diameter $d$ (blue lines), and the barrier height $V_B$ (red lines). The leftmost points, for zero variation, are the values for the ordered channel which includes an $8 \times 4$ hexagonal array of NIs with fixed $d = 3$ nm and $V_B = 0.05$ eV. We consider variations up to $30\%$ in the parameters (and use averaged data from at least 10 simulations for each point). It can be seen that variation in barrier height has negligible effect on $G$ and $S$, and due to the adverse interdependence of $G$ and $S$, the minor effect that is seen cancels out leaving no significant change in the PF even at $30\%$ variation. Likewise, variation in the NI diameter, although slightly more consequential than barrier height, shows little impact on the thermoelectric coefficients. [Interestingly a small positive effect on the PF is seen, although this is probably due to the statistical variation in diameter leading to a small increase in overall density, increasing the small energy filtering effect of the barriers].
\begin{figure}[t]
\vspace*{-0in}
\includegraphics[width=7.4cm,height=13cm]{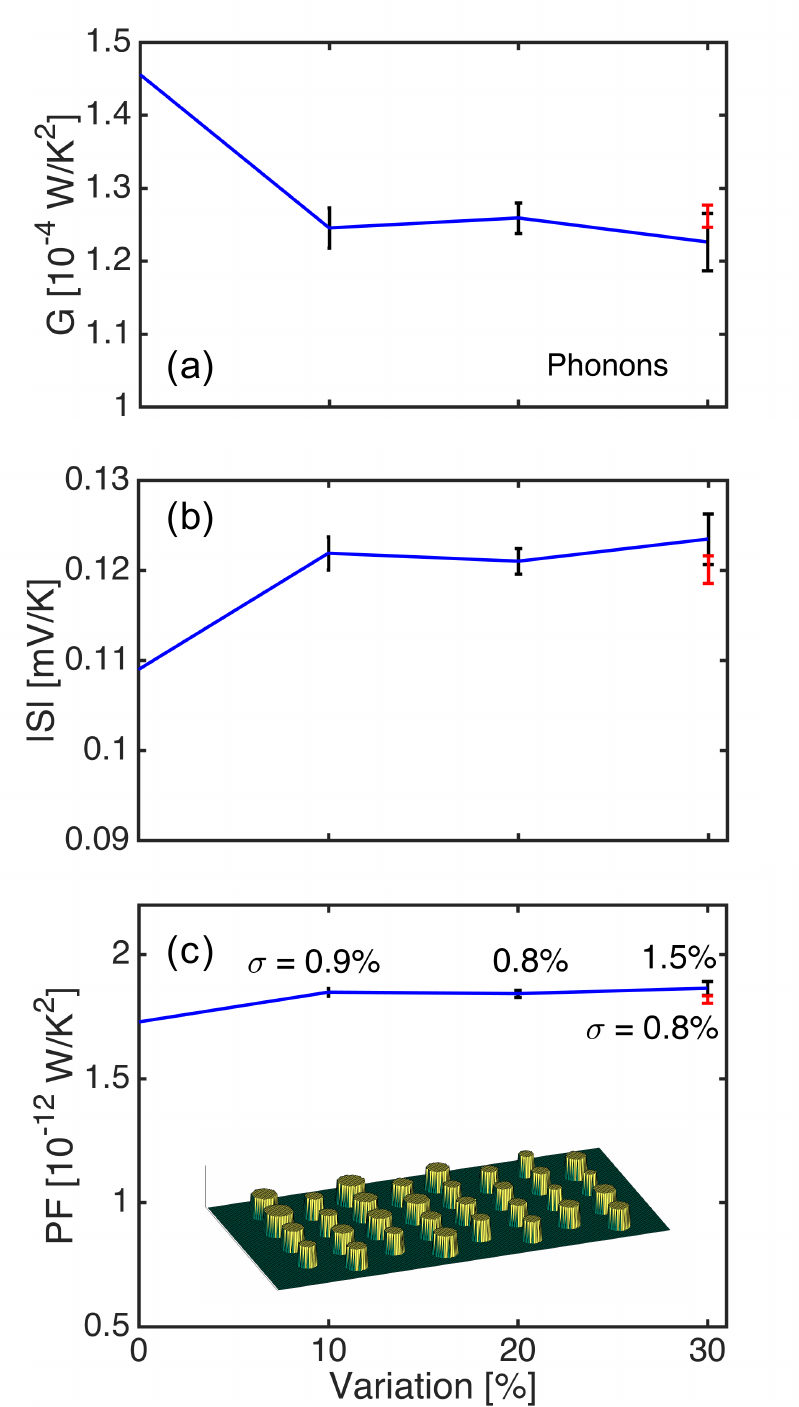}
\vspace*{0.05cm} \caption{\label{fig4} (Colour online)
The effect of variations in the diameter (blue lines) and barrier height (red points) on the thermoelectric coefficients. (a) The conductance, (b) the Seebeck coefficient, (c) the PF. In the inset of (c) we show an example geometry with a $30\%$ variation in diameter. In (c) we also show the standard deviations in our values of the PF.
}
\end{figure}

The next investigation we perform is to explore the effect of variations in the geometry of the channel. We again simulate a ballistic channel with a total of $32$ cylindrical barriers of height $V_B = 0.05$ eV and diameter $d = 3$ nm. In this case each of the NIs is randomly placed in the channel (as shown in Fig.~2(c)) rather than in an ordered hexagonal fashion. The NIs are treated as a region in the matrix material with a different band energy, i.e. from the electronic point of view, these are regions which built potential barriers in the matrix material. One could be more flexible as to define a different effective mass as well in those regions, as well as built-in potentials due to charge fluctuations which could lead to Schottky barriers, etc., but this is beyond the scope of this work. Other than that, NEGF is a real space technique, where transport is described quantum mechanically and the details of the geometry, accurately. To be able to treat real materials, one could employ details of the bandstructures (effective masses, degeneracies, mean-free-paths, phonon energies), possibly extracted from Density Functional Theory calculations, but again this is beyond the scope of this work. 

In Fig.~5 we show the histograms of the conductance $G$ (Fig.~5(a)), Seebeck coefficient $S$ (Fig.~5(b)), and PF $GS^2$ (Fig.~5(c)) of $40$ random geometries and a Gaussian fit calculated from the results. It can be seen that, although the standard deviation values are slightly higher than the previous cases considered, most values fall within just a few percent of the mean.

In order to achieve a complete picture of the transport through disordered nanocomposites, we now consider the effect of variations on a channel in the acoustic phonon scattering regime. In Fig.~6 we show the thermoelectric conductance $G$ (Fig.~6(a)), Seebeck coefficient $S$ (Fig.~6(b)), and PF $GS^2$ (Fig.~6(c)) of a channel with an $8 \times 4$ hexagonal array of NIs as two different parameters are varied: the barrier diameter $d$ (blue lines), and the barrier height $V_B$ (red points). Similarly to the ballistic case (compare with Fig.~5) variation in the diameters of the NIs produces only small changes in $G$ and $S$ which cancel out in the PF leaving it relatively unchanged. In the same way, small variations in the barrier height have no significant impact on the PF. Here we only simulated variations at the $30\%$ level for the barriers since it will provide the largest variation in the PF of the percentages that we have considered.

It is of interest at this point to compare our above results to previous works on variations in superlattices \cite{thesberg15, thesberg2}. It was previously shown that the thermoelectric transport in superlattices is highly sensitive to variations in the barrier heights. This is because in superlattices each electron must pass through each individual barrier region, providing a strong energy filtering effect. The height of each barrier degrades the conductivity exponentially, meaning variations away from the optimal structure can have a significant impact on the power factor. In the NI case, however, electrons can flow around the NIs. In fact, most of the current of the flow is through the matrix material \cite{fos1} which is not affected by variations in the barrier height. Variations in the NI structure therefore do not have a large effect on the electron transport and the power factor is similarly unaffected. 

\textit{Approximations and simplications:} We would like to comment on certain approximations/simplifications that we have made in this work. First, even though a local potential barrier is a legitimate way to model a NI to first order, in real materials there may be deformations in the vicinity of a NI, which can produce strain fields that lead to modified effective mass and band details. In addition, we ignored self-consistent charging effects that could alter the shape of the barrier. Further, we considered only ballistic transport and phonon scattering, and neglected ionized impurity scattering, which could suppress the PFs quantitatively \cite{neo13}, but qualitatively would not change the results on the effects of variations. 

Finally, we mention that the treatment of phonons in this paper is simplified and is adjusted to provide a certain reasonable mean-free-path (already quite complicated computationally within a fully quantum mechanical formalism, however). The details of the phonon spectrum are even neglected, however, in any case in this study we are after qualitative conclusions for the power factor in the presence of variations, which seems to be minimal anyway and even smaller when phonon scattering gets stronger. Also, we do not consider a specific material, but keep the study as generic as possible, focusing qualitatively on the geometrical influences on transport. A more detailed el-ph model is possible, but computationally expensive, possibly prohibitive for the size of structures we consider here. Previous works \cite{luis09} have considered full phonon dispersions to construct the self-energies that enter the Green’s function, but those studies were limited to nanowire channels of only up to 3nm in diameter.

We also point out that transport in the type of structures that we consider here is not much dependent on the effective mass. Our previous work \cite{fos1} revealed that changing the effective mass does not affect the influence of variability. When it comes to extending this study to realistic, complex bandstructure materials, atomistic techniques (tight-binding, DFT) can be used to provide effective parameters, since coupling atomistic methods to NEGF is accompanied by an enormous computational cost and limit the size of the structure that we simulate.

\section{Conclusions} 

In conclusion, using the fully quantum mechanical non-equilibrium Green’s function method, we investigated the effect of random variations in the parameters of a $2D$ nanocomposite channel on the thermoelectric coefficients: conductance, Seebeck coefficient, and power factor. We showed that, unlike superlattices, materials with NIs are robust to variations in the barrier height. We also showed this robustness holds for variations in the NI diameter and NI geometry. Our findings suggest a design regime for nanocomposites that should provide power factor robustness while achieving reductions in the lattice thermal conductivity. In particular, we showed a design regime for which the density and geometry can be optimized for maximum phonon scattering and maximum reduction in thermal conductivity while preserving the power factor, producing high $ZT$.

{\bf Acknowledgments} This work has received funding from the European Research Council (ERC) under the European Union's Horizon 2020 Research and Innovation Programme (Grant Agreement No. 678763).



\end{document}